\documentclass[article]{jss}

\usepackage[latin1]{inputenc}
\usepackage[T1]{fontenc}
\usepackage{amssymb}
\usepackage{url}
\usepackage{amsmath}
\usepackage{algorithm}
\usepackage{algorithmic}

\newcommand{\tlaw}{\widetilde{\law}}
\usepackage{./StefanTexCommandsV2}

\author{Julie Josse\\Ecole Polytechnique - INRIA \And 
     Sylvain Sardy    \\ University of Geneva \\
\And Stefan Wager\\ Stanford University	 \\
				}
\title{\pkg{denoiseR}: A Package for Regularized \\ Low-Rank Matrix Estimation}

\Plainauthor{Julie Josse, Sylvain Sardy, Stefan Wager} %% comma-separated
\Plaintitle{\pkg{denoiseR} a package for low rank matrix approximation} %% without formatting
\Shorttitle{\pkg{denoiseR}  } %% a short title (if necessary)

\Abstract{
We introduce \pkg{denoiseR}, an \proglang{R} package that provides a unified
implementation of several state-of-the-art proposals for regularized low rank matrix estimation,
along with automatic selection of the regularization parameters.
We also extend these methods to allow for missing values.
The regularization schemes discussed in this paper are built around singular-value shrinkage and
bootstrap-based stability arguments.
We illustrate how to use out package by applying it to several real and simulated datasets, and
highlight strengths and weaknesses of the different implemented methods.
}

\Address{
  Julie Josse\\
  Department of Statistics\\
  Agrocampus Ouest/ INRIA / Polytechnique\\
  E-mail: \email{julie.josse@polytechnique.edu}\\
  URL: \url{http://julie.josse.com}
}

\begin{document}

\section{Introduction}

Consider the model  where a data matrix $X$ with $n$ rows and $p$ columns is generated
 from some distribution $\law(\mu)$ with $\EE[\mu]{X} = \mu$:
 \begin{equation} \label{mod:lr}
 X \in \RR^{n \times p} \sim \law(\mu) \mbox{ with } \mu \mbox{ of low rank } k.
 \end{equation} 
The statistical aim is to recover the signal $\mu$ from the noisy data.
The low-rank assumption underlying model \eqref{mod:lr} has become very popular in recent
years, and arises naturally in several different settings \citep{Udell17}.
It gained traction in the machine learning community as a powerful way to address the problem
of recommender systems, as exemplified by the famous Netflix challenge \citep{Netflix}.

The classical approach to this problem is to estimate the signal $\mu$ as the best rank-$k$ approximation
to $X$, for some adaptively chosen $k$:
\begin{equation}\label{eq:ls}
\argmin_\mu \left\{\Norm{X - \mu}_2^2 : \rank{\mu} \leq k\right\}.
\end{equation} 
The solution is the truncated singular value decomposition (SVD)
of the matrix $X=UDV^\top$ at the order $k$ \citep{Eckart:1936}, namely
\begin{equation} \label{eq:tsvd}
\hmu_k = \sum_{l = 1}^{k} d_l \, u_l \,  v_l^\top,
\end{equation}
where $d_l$ are the singular values organized in decreasing order.

In recent years, however, there as been considerable interest in procedures that improve
over the truncated SVD by further regularizing \smash{$\hmu$}.
These proposals include matrix soft-thresholding \citep{CandesSURE:2013},
adaptive trace norm regularization \citep{sardy15},
bootstrap-based regularized autoencoding for count data and other exponential family noise models \citep{JosseWager},
and estimators motivated by asymptotic expansions \citep{Verbanck:RegPCA:2013, OS:2014};
see Section~\ref{sec:lrs} for a more detailed description.
We implement all these methods in the \proglang{R} package \pkg{denoiseR}. 
We give a particular attention to providing sensible choices of default parameters, automatic selection of the regularization parameters and ways to estimate the noise variance.

The package \pkg{denoiseR}, available on CRAN, is the first to implement  low rank matrix estimation methods in the \proglang{R} language.
We describe in Section~\ref{sec:implementation} the main functionalities of \pkg{denoiseR} and give guidelines on the predilection regimes of each method.
% \fbox{We also discuss the sensitivity of the methods to the regularization parameters?}
In the remaining sections, we go beyond published works as follows. 
In Section~\ref{sec:comp}, we extend all methods to the incomplete case, and
 tackle the challenging task of selecting the regularization parameter(s) when missing values are present by extending the Stein unbiased risk estimation.
As an aside, we propose a new highly competitive method to impute data and to  complete count data. % which offers new perspectives.
Finally, Section \ref{sec:data} carries out experiments and illustrates the potential of the methods to denoise data sets from different fields. More precisely, an unsupervised clustering is performed on a microarray data, correspondence analysis  is used to visualize  a documents-words data of the inaugural speeches of the U.S.~Presidents,
and matrix completion is applied on the ``journal impact factors'' data. 

Software for regularized singular value decomposition is partially available.
Soft-thresholding estimator with Stein unbiased risk estimation is implemented in \proglang{MATLAB} as a standalone file that is available at \url{https://statweb.stanford.edu/~candes/SURE/}.
The asymptotic estimators of \citet{Raj} and \citet{OS:2014} is available as a \proglang{MATLAB} software library which can be downloaded in \url{https://purl.stanford.edu/kv623gt2817} and that includes
a function to calculate the optimal singular value shrinkage with respect to  the Frobenius operator and nuclear norm losses, both in known or unknown noise level. 
Many \proglang{R} packages implement versions of classical truncated SVD
(online, fast, etc.), but as far as we know, no method for regularized low rank matrix estimation is available, which is the aim of \pkg{denoiseR}.
One should mention related packages on covariance matrix estimation with shrinkage strategies such as \pkg{covmat}, \citep{covmat} \pkg{corpcor} \citep{corpcor} or \pkg{nlshrink} \citep{nlshrink}.
Packages on missing values are detailed in Section~\ref{sec:comp}.

% compared to competitors \citep{Trevor15} with %simulations.
%To compare several matrix completion methods,
%we perform a simulation study  such as the one suggested in \citet{Trevor15}  and implemented in the  %\proglang{R} package \pkg{softImpute}  \citep{softimpute}.

%%%%%%%%%%%%%%%%%%%%%%%%%%%%%%%%%%%%%%%%%%%%%%%%%%%%%%%
\section[]{Methods for low-rank matrix estimation} \label{sec:lrs}

% In this section, we introduce low-rank matrix estimators that have been defined in a finite sample framework, in an asymptotic framework and using a bootstrap approach.  

\subsection[]{Singular values shrinkage} \label{sec:finite}

We start by considering direct extensions of the truncated SVD estimator \eqref{eq:tsvd} using
singular value shrinkage.
Perhaps the best known estimator of this type arises via the nuclear norm regularized method
studied by \citet{Cai08}:
\begin{equation*} \label{eq:softc}
\argmin_\mu  \left\{\frac{1}{2}\Norm{ X- \mu}_2^2 +  \lambda {\Norm{\mu }_{*}}\right\},
\end{equation*}
where ${\Norm{\mu }_{*}}$ is the nuclear norm of the matrix $\mu$.
Algorithmically, this method is equivalent to soft-thresholding singular values in \eqref{eq:tsvd}
(see \citet{Dono94b} for definitions of hard and soft thresholding functions): 
\begin{equation} \label{eq:soft}
\hmu_k = \sum_{l = 1}^{\min\{n, \, p\}} d_l \max\left\{1- {\lambda}/{d_l},0\right\} \, u_l \,  v_l^\top;
\end{equation}
In other words, singular values smaller than a quantity $\lambda$ are set to zero and the others are
shrunk towards zero by an amount $\lambda$ (in contrast to hard thresholding, where singular values
are either unmodified or then set to zero).
The number of non-zero singular values provides an estimation for the rank $k$.
Many studies \citep{shabalin2013reconstruction,OS:2014,sardy15, JosseWager} showed both analytically and with simulations that
the soft-thresholding estimator gives small mean squared error (MSE) to recover $\mu$
with low signal-to-noise ratio (SNR) but struggles in other regimes.

For more flexibility, the adaptive trace norm estimator (ATN) of \citet{sardy15} uses two regularization parameters $(\lambda, \gamma)$ to threshold and shrink the singular values: 
\begin{equation} \label{eq:sard}
 d_l \max\left(1-\frac{\lambda^\gamma}{d_l^\gamma},0\right).
\end{equation}
This estimator denoted $\hat \mu_{(\lambda, \gamma)}$ is the closed form solution to
\begin{equation*}
\argmin_\mu  \left\{\frac{1}{2} \Norm{ X- \mu}_2^2 +  \lambda^\gamma {\Norm{\mu }_{*,w}}\right\}, 
\end{equation*}
where ${\Norm{\mu }_{*,w}} =\sum_{l=1}^{\min(n,p)} \omega_l d_l $ is a weighted nuclear norm with $\omega_l=1/d_l^{\gamma-1}$.
ATN parametrizes a rich family of estimators and includes  \eqref{eq:tsvd} and \eqref{eq:soft} as special cases.
As an insight of its good behavior,   the smallest singular values responsible for instability in \eqref{eq:sard} are more shrunk in comparison with the largest ones when $\gamma>1$.
% which is satisfactory since the smallest singular values are responsible for instability. 

The parameters $(\lambda, \gamma)$ can be selected with  cross-validation.
However, in the context of a Gaussian specialization of the low-rank model \eqref{mod:lr},
\begin{equation} \label{mod:gauss}
X= \mu + \varepsilon \quad \with \quad \varepsilon_{ij} \simiid \nn\p{0, \sigma^2}  \mbox{ and } \mu \mbox{ of low rank } k,
\end{equation}
more computationally efficient tuning is possible using
Stein  unbiased estimate of the risk $\mathbb{E} \|\mu  - \hat \mu_{(\lambda, \gamma)}\|^2_2$ \citep{Stein:1981}.
\citet{sardy15} extend the results of \citet{CandesSURE:2013} and propose the risk estimate
\begin{equation}
\label{eq:SURE}
{\rm SURE} (\lambda, \gamma) = - np\sigma^2 +
\sum_{l=1}^{\min(n,p)}d_l^2 \min\left(\frac{\lambda^{2\gamma}}{d_l^{2\gamma}}, 1\right) + 2 \sigma^2 {\rm div}(\hat \mu_{\lambda,\gamma} ),
\end{equation}
where the second term corresponds to the residuals sum of squares (RSS) whereas the last one is the divergence defined as
\begin{eqnarray*}
{\rm div}(\hat \mu_{\lambda,\gamma})&=&\sum_{l=1}^{\min(n,p)} \left(1+(\gamma-1) \frac{\lambda^\gamma}{d_l^\gamma}\right) \cdot 1\left(d_l \geq \lambda\right)
+ |n-p| \max(1-\frac{\lambda^\gamma}{d_l^\gamma},0)\\
&&+ 2 \sum_{t \neq l,t=1}^{\min(n,p)} \frac{d_l^2 \max(1-\frac{\lambda^\gamma}{d_l^\gamma},0) }{d_l^2-d_t^2}.
\end{eqnarray*}
The expectation of the divergence corresponds to the degrees of freedom.
Having a closed form expression for the divergence is computationnally convenient.
Without it, the divergence could be approximated by finite differences, which is our approach with missing values in~\eqref{eq:sureatnna}.

A limitation of SURE is that it requires knowledge of the noise scale $\sigma^2$.
Inspired by generalized cross-validation \citep{CravenWahba79}, \citet{sardy15} derived generalized SURE
\begin{equation}
\label{eq:GSURE}
{\rm GSURE} (\lambda, \gamma)  =  \frac { \mbox{RSS}}{(1-{\rm div}(\hat \mu_{\tau,\gamma})/(np))^2},
\end{equation}
which does not require knowledge of $\sigma^2$.

The parameters $(\lambda, \gamma)$ are then estimated by minimizing (G)SURE. 
However, minimizing (G)SURE can prove to be difficult and unstable.
The difficulty is due to the fact that (G)SURE is a piecewise smooth surface with occasionnal jumps as a function of $\lambda$, and the higher $\gamma$, the higher the jumps.
The unstability of the (G)SURE selection is due to the fact that, although unbiased, the risk estimate has some variance.
Consequently, in situation of low signal-to-noise ratio, the global minimum may 
occur at a value of $(\lambda, \gamma)$ that has poor estimation properties for the low rank matrix
$\hat \mu_{\lambda, \gamma}$.  \citet{SardySBITE2012} studied the smoothness and the minimization of SURE in regression, and \citet{sardy15} compared a SURE surface to the $\ell_2$-loss surface in low rank matrix estimation.

Of course, minimizing over $\lambda$ for $\gamma=1$ (i.e., soft-thresholding)
is easier than minimizing over $(\lambda, \gamma)$.
In addition, when estimation of $\sigma^2$ is poor, SURE may lead to a poor selection of the regularisation parameter(s).

Nevertheless, the ATN estimator  \eqref{eq:sard} has shown excellent recovery properties in experiments \citep{sardy15, JosseWager},
in particular in comparison with the soft-thresholding estimator~\eqref{eq:soft}. This is particularly true when the signal-to-noise ratio is moderate or high,
in which case the minimum of (G)SURE over $(\lambda, \gamma)$ is close to the minimum of the (unknown) $\ell_2$-loss and therefore leads to good estimation with ATN.
For good rank recovery, \citet{sardy15} employed the quantile universal threshold (QUT)  to select $\lambda$ \citep{Sardy16}.
% Indeed, $\lambda$ controls the rank of the solution via the number of non-nul singular values.
The rationale of QUT is to select the threshold $\lambda^{\rm QUT}$ at the bulk edge of what a threshold should be to reconstruct the correct model with high probability under
the null hypothesis that $\mu=O$ (the $n\times p$ matrix with zeros for entries).
Then they minimize  GSURE($\lambda^{\rm QUT}, \gamma$) over~$\gamma$.
% More precisely, they generate 1000 times data under the null hypothesis of no signal, $\mu= 0$ and use the $(1-\alpha)$-quantile of the distribution of the largest empirical singular value as a threshold.
% With $\alpha$ tending to zero with the sample size, null rank estimation is guaranteed with probability tending to one under the null hypothesis. 

%In this section, we provide details on how to use \pkg{denoiseR} to employ the estimators listed in the previous sections.

\subsection[]{Asymptotically optimal shrinkage} \label{sec:asympt}

In certain asymptotic regimes, it is possible to go beyond the results presented above, and to derive
optimal singular value shrinkage estimators that minimize expected loss.
First, when both the number of rows ($n = n_p$) and columns ($p$) tend to infinity while the rank of the matrix stays fixed, \citet{shabalin2013reconstruction} and \citet{OS:2014} consider asymptotics motivated by random matrix theory
and show that the estimator the closest to the true signal $\mu$ in term of MSE has the following form,
\begin{equation} \label{eq:shaba}
 \frac{1}{d_l} \sqrt{\left(d_l^2 - (\beta -1)n\sigma^2 \right)^2 - 4 \beta n\sigma^4} \cdot 1\left(l\geq (1+\sqrt{\beta})n\sigma^2\right), 
\end{equation}
where $n_p/p \rightarrow \beta$, $0 < \beta \leq 1$, and
$1()$ is the indicator function (note that the restriction $n_p \leq p$ is without loss of generalization,
since we can always apply this formula to the transpose of the matrix).
If the noise variance is unknown, the authors suggested
\begin{equation}\label{eq:sigma_hatdono}
\hat \sigma= \frac{d_{med}}{\sqrt{n \mu_{\beta}}},
\end{equation}
where $d_{med}$ is the median of the singular values of $X$ and $\mu_{\beta}$ is the median of the Marcenko-Pastur distribution.
\citet{OS:2014} also provided optimal shrinkers using other losses than the Frobenius one, namely the Operator and Nuclear losses. 

In another asymptotic framework, considering $n$ and $p$ as fixed, but letting the noise variance $\sigma^2$ tend to zero, \citet{Verbanck:RegPCA:2013}
shows that the MSE-minimizing estimator
has the following form:
\begin{eqnarray}
\label{eq:verb}
     d_l \left(\frac{d_l^2-\sigma^2}{d_l^2}\right) \cdot 1(l\leq k).
\end{eqnarray}
Each singular value is multiplied by a quantity which can be seen as the ratio of the signal variance over the total variance. 
This estimator corresponds to a truncated version of the one suggested in \citet{Efron72}.
If the noise variance is unknown, the authors suggested 
\begin{equation} \label{eq:sigma_hat}
\hat \sigma^2= \frac{\Norm{X- \sum_{l=1}^ku_ld_lv_l}_2^2 }{np-nk-kp+k^2},
\end{equation}
which corresponds to the residual sum of squares divided by the number of observations minus number of estimated parameters. 
The variance estimate requires a value for the rank $k$, for instance estimated by cross-validation \citep{Josse11b}. This latter method has shown excellent empirical performances. 
%This specific asymptotic framework is in agreement with data where both the rows and columns can be %considered as fixed while the randomness
%comes from the measurement errors. 
%Such data are common in sensory analysis for instance where rows are specific products such as beers, %columns are descriptors such as bitterness, sweetness, etc and each cell corresponds to the average scores %given by a set of panelists for one product and one descriptor.
%The asymptotic corresponds to more judges being sampled. 

\subsection{Bootstrap-based regularization for exponential family noise models} \label{sec:boot}

All previous estimators started from the point of view of the singular value decomposition.
\citet{JosseWager} suggested an alternative regularization based on the bootstrap.
Given an observed matrix $X \sim \law (\mu)$, the optimal rank $k$ linear estimator for $\mu$ would be
\begin{equation} \label{eq:oracle}
\hmu^{(k)^*} = XB^{(k)^*}  \where  B^{(k)*} = \argmin_B \left\{\EE[X \sim \law(\mu) ]{\Norm{\mu - XB}_2^2} : \rank{B} \leq k\right\}.
\end{equation}
This estimator, however, is infeasible, as it involves computing an integral over an unknown data distribution.

To get around this issue, \citet{JosseWager} propose replacing the draws $X \sim \law\p{\mu}$ from the unknown data-generating distribution with bootstrap draws $\tX \sim \tlaw_\delta\p{X}$, resulting in an estimator they call the \emph{stable autoencoder},
\begin{eqnarray} 
\label{eq:noise}
\hmu^{(k)} = X\hB^{(k)}, \ \ \hB^{(k)} &=& \argmin_B \left\{\EE[\tX \sim \tlaw_\delta(X)]{\Norm{X - \tX B}_2^2} : \rank{B}  \leq k\right\}.
\end{eqnarray}
With Gaussian data as in \eqref{mod:gauss}, \citet{JosseWager} advocate Gaussian parametric bootstrap $\tX_{ij} \sim \nn\p{X_{ij}, \, \delta / (1 - \delta) \, \sigma^2}$, while with Poisson data $X_{ij} \sim \text{Poisson}\p{\mu_{ij}}$, they recommend binomial thinning
$$\tX_{ij} \sim \frac{1}{1 - \delta} \, \text{Binomial}\p{X_{ij}, \, 1 - \delta}. $$
Here $\delta \in (0, \, 1)$ is a regularization parameter controlling the amount of bootstrap noise.
Both of these proposals are special cases of a L\'evy bootstrap procedure that can be defined for any exponential family distribution \citep{wager2016data}.

Given a parametric bootstrap scheme, the problem \eqref{eq:noise} can equivalently be written as,
$$\hB^{(k)}= \argmin_B \left\{\Norm{X - XB}_2^2 + \Norm{S^{\frac{1}{2}}B}_2^2   : \rank{B} \leq k \right\}  \quad {\rm with} \quad
S_{jj} = \sum_{i = 1}^n \Var[\tX \sim \law\p{X}]{\tX_{ij}}$$ and the solution is 
\begin{eqnarray}
\label{eq:sacount}
 \hat \mu = X(X^{\top}X + S)^{-1} X^{\top} X.
\end{eqnarray}

When, the noise is Gaussian as in model \eqref{mod:gauss}, $S$ is equal to a diagonal matrix with elements $n\sigma^2$ and \eqref{eq:sacount} can also be written as a classical singular-value shrinkage estimator:
\begin{eqnarray*} 
\hmu^{(k)}_\lambda &=& \sum_{l = 1}^k u_l \, \frac{d_l}{1 + \lambda/d_l^2} \, v_l^\top \quad \with \quad \lambda = \frac{\delta}{1 - \delta} n\sigma^2.
\end{eqnarray*}
If instead  we have count data and consider a Binomial model,
$S$ is a diagonal matrix with row-sums of $X$ multiplied by ${\delta}/\p{1-\delta}$.
Due to the non-isotropic noise, the new estimator $\hmu_{\delta}$ \eqref{eq:sacount} does not reduce to singular value shrinkage: its singular vectors are  also modified. This characteristic is unique and implies that the estimator will be better than the competitors that only shrink the singular values to recover the signal for models such as $X_{ij} \sim \text{Poisson}\p{\mu_{ij}}$. 
The complexity of estimator $\hmu_{\delta}$ is determined via $\delta$ that we estimate by cross-validation.
%for a fixed $\delta$, remove  a set of observed cells $X_{ij}$ and  predict them with $\hmu_{\delta}$ obtained from the data set
%that excludes these cells; then, the prediction error $(\hat \mu_{{\delta}_{ij}}^{-ij} - X_{ij})^2$ is computed and the operation repeated for $\delta$ varying from 0 to~1.
%The selected value of $\delta$ minimizes the mean squared error of prediction (MSEP).
This procedure requires to get $\hmu_{\delta}$ from an incomplete data set, which will be described in Section~\ref{sec:comp}.

In a second step, \citet{JosseWager} showed that the performance of stable autoencoding can considerably be improved by iterating the procedure until reaching a fixed point of the proposed denoising scheme.  Specifically, they iterate the procedure by replacing $X$ with the low rank estimate  $\hmu = X \hB$ obtained from the optimal $B$,
and by solving  $\eqref{eq:noise}$ with $\hat B = (\hat \mu^{\top}\hat \mu + S)^{-1} \hat \mu^{\top} \hat \mu$ as described in Algorithm \ref{alg:iter}. 

%\spacingset{\SPACESMALL}
\begin{algorithm}[t]
\caption{Low-rank matrix estimation via iterated stable autoencoding.}
\label{alg:iter}
\begin{algorithmic}
\STATE $\hmu \gets X$
\STATE $S_{jj} \gets \sum_{i = 1}^n \Var[\tX \sim \tlaw_\delta\p{X}]{\tX_{ij}}$ for all $j = 1, \, ..., \, p$
\WHILE{algorithm has not converged}
  \STATE $\hB \gets \p{\hmu^\top \, \hmu + S}^{-1} \hmu^\top \, \hmu$
  \STATE $\hmu \gets X \, \hB$
\ENDWHILE
\end{algorithmic}
\end{algorithm}
%\spacingset{\SPACEBIG}

One advantage of the \emph{iterated stable autoencoder} (ISA) is that it automatically produces low rank estimates $\hat{\mu}$;
thus, the practitioner only needs to specify a single regularization parameter $\delta$, instead of having to choose both $k$ and $\delta$ with the original stable autoencoder \eqref{eq:noise}.
The procedure requires knowledge of $\sigma$ for Gaussian models.
%This estimator can substantially outperform its competitors with non-Gaussian noise models.

Finally, \citet{JosseWager} extended their results and also used ISA to regularize correspondence analysis (CA) \citep{green1984ca, green2007ca}. 
CA is a powerful method to visualize contingency tables and consists in applying an SVD on the data transformation
\begin{equation}
\label{eq:ca}
M = R^{-\frac12} \p{X - \frac{1}{N} rc^\top} C^{-\frac12},
\end{equation}
where $ R = \diag\p{r}$, $C = \diag\p{c}$,
$N$ is the the total number of counts, and $r$ and $c$ are vectors containing the row and column sums of $X$. This approach will be illustrated Section \ref{sec:presi}.

\section[]{Low rank matrix estimation with denoiseR} \label{sec:implementation}

\subsection[]{Implemented methods} \label{}

\pkg{denoiseR} is designed to estimate a low rank signal with the estimators described in the previous sections.
Table \ref{table:recap} lists the methods (for Gaussian noise), their \proglang{R} names, some options, the selection rule for their regularization parameter(s) and
 their regime of predilection.
 Additionally, the functions   \code{estim_sigma} and 
\code{estim_delta} are provided to respectively estimate the noise variance and the bootstrap noise for ISA. 

\begin{table}[t]
\scriptsize
\centerline{
\begin{tabular}{|r|r|r|r|r|r|r|}
\hline
method& function & option &  regularization parameter& $\sigma$ required & setting  \\
\hline
Soft threshold \eqref{eq:soft} &  \code{adashrink} & gamma.seq=1   & $\lambda$ with SURE or QUT &  yes  & low SNR\\ 
Soft threshold \eqref{eq:soft} &  \code{adashrink} &gamma.seq=1&  $\lambda$ with GSURE & no & low SNR\\ 
ATN \eqref{eq:sard}&  \code{adashrink} &&  $(\lambda, \gamma)$ with SURE or QUT & yes & medium,high SNR \\
ATN \eqref{eq:sard} &  \code{adashrink} & & $(\lambda, \gamma)$ with GSURE & no & good overall \\ 
Asympt  \eqref{eq:shaba}  &\code{optishrink}& method=ASYMP  & & yes &  $n$, $p$ $\rightarrow \infty$ \\
Low-noise \eqref{eq:verb} & \code{optishrink} & method=LN & & yes &  $\sigma \rightarrow 0$ \\
ISA (Algo \ref{alg:iter}) & \code{ISA} &   & $\delta$ with  CV& no &  moderate SNR\\ 
\hline
\end{tabular}
}
\protect\caption{Methods available in \pkg{denoiseR} to denoise continuous data. \label{table:recap}}
\normalsize
\end{table}

%\subsection[]{Recommendation} \label{}

%using the three functions \code{adashrink} for the ATN estimator \eqref{eq:sard}
%and \code{optishrink} for the two asymptotic estimators \eqref{eq:shaba}  and \eqref{eq:verb},
%along with \code{estim_sigma} to estimate the noise variance.

The different methods have been extensively compared using simulations in \citet{sardy15} and \citet{JosseWager}.
The simulations highlighted that the proposed methods have different strengths and weaknesses.
The soft-thresholding estimator \eqref{eq:soft} behaves well in low SNR settings, but struggles in other regimes.
The other estimators,  ISA (Algorithm~\ref{alg:iter}), Asympt \eqref{eq:shaba}, Low-noise \eqref{eq:verb}, with non-linear singular-value
shrinkage functions are more flexible and perform well expect when the SNR is low.
The estimators Asympt and Low-noise provide good recovery in their asymptotic regimes.
With two regularization parameters $\lambda$ and~$\gamma$, the ATN estimator \eqref{eq:sard} is flexible and can adapt well to the signal-to-noise ratio.
Often, the two-dimensional SURE criterion identifies a minimum that leads to good estimation. On occasion, the SURE surface is a rather erratic function of $\lambda$ and $\gamma$,
causing the estimator to perform poorly, in particular in low signal-to-noise ratio. Estimation of the noise variance has necessarely an impact on the results and suggested methods accurately estimate the noise variance in their respective regime but may struggle in others. 
Nevertheless, the ATN estimator,  with its procedure to select the regularization parameters,  performs best overall.

Based on these experiments, we recommend the following strategy for regularized low rank matrix estimation. 
If one of the asymptotic regimes discussed in Section \ref{sec:asympt} appears plausible,
the user should use \code{optishrink} with either the \code{ASYMPT} or \code{LN} option.
In other settings, for Gaussian noise,  if the noise variance is unknown,   ATN with GSURE should be used, as it does not require an estimate of $\sigma$. % in the first step of the calculations.
If GSURE encounters any difficulties, we suggest first estimating  the variance (with the method that is the most plausible given the data at hand) and then using ATN with SURE. 
ISA is recommended for non-Gaussian noises. 
Finally,  ATN with $\lambda_{\mbox{QUT}}$  gives good indication for rank estimation.
 ISA also tends to estimate the rank accurately except when the signal is nearly indistinguishable from the noise while the soft-thresholding  tends to over-estimate the rank.

The next subsection provides the arguments of the different functions. We start with ATN and GSURE as recommended and then describe how to perform ATN with SURE. Soft-thresholding is included in the package as it is a classical method.  

%%%%%%%%%%%%%%%%%%%%%%%%%%%%%%%%%%%%%%%%%%%%%%%%%%%%%%%%%%%%%%%%%%%%%%%%%%%%%%%%
\subsection[]{Using denoiseR with Gaussian noise} \label{sec:codelr}

After installing the \pkg{denoiseR} package, we load it.
First, we generate a data set of size $n \times p$ of rank $k$ according to model \eqref{mod:gauss} with the function \code{LRsim}. 
The amount of noise is defined by the argument \code{SNR}$=1/ (\sigma \sqrt{np})$.
This function returns the simulated data in~\code{X}, the signal in~\code{mu} and the standard deviation of the noise in~\code{sigma}.

\begin{CodeInput}
R> library("denoiseR")
R> Xsim <- LRsim(n = 200, p = 500, k = 10, SNR = 4)
\end{CodeInput}

Then, we denoise the data with ATN \eqref{eq:sard} using the function \code{adashrink} as follows:
\begin{CodeInput}
R> ada.gsure <- adashrink(Xsim$X)
R> muhat <- ada.gsure$mu.hat
\end{CodeInput}
\begin{CodeOutput}
R> ada.gsure$nb.eigen
[1] 10
\end{CodeOutput}

By default, the regularization parameters ($\lambda, \gamma$) are selected with GSURE \eqref{eq:GSURE}.
The function outputs the estimator  in \code{mu.hat} which is a matrix of size $n \times p$,
the  number of non-zero singular values in \code{nb.eigen} (here 10), the optimal~$\gamma$ in \code{gamma},
the optimal~$\lambda$ in \code{lambda}, and the SVD of~$\hat \mu_{(\lambda, \gamma)}$ in \code{low.rank} and  \code{singval}.

The regularization parameters can also be selected with the other strategies by specifying the argument \code{method} to \code{SURE} or \code{QUT}.
The first one is recommended to estimate the signal and the second to estimate the rank of $\mu$. % via the number of non-zero singular values.
As detailed in Section \ref{sec:finite}, both strategies require the variance of the noise (as opposed to GSURE).

%R> ada.sure <- adashrink(Xsim$X, method = "SURE", sigma = Xsim$sigma)
\begin{CodeInput}
R> ada.sure <- adashrink(Xsim$X, method = "SURE")
\end{CodeInput}
\begin{CodeOutput}
Warning message:
In adashrink(Xsim$X, method = "SURE") : sigma estimated by MAD: 0.000805
\end{CodeOutput}
 %In the first line,
%we provided the true value  (\code{Xsim$sigma}). If nothing is specified, it is estimated 
If $\sigma^2$ is unknown, \code{adashrink} calls by default the function \code{estim\_sigma} with the argument \code{method = "MAD"}
to estimate it with \eqref{eq:sigma_hatdono}. 
If low-noise   seems more plausible, estimator \eqref{eq:sigma_hat} can be obtained by specifying the argument \code{method = "LN"},
which requires knowing the rank~$k$. Otherwise, $k$ is estimated by default by cross-validation as implemented in the \code{estim\_ncp} function of the \pkg{FactoMineR} package  \citep{Facto}:
 
\begin{CodeInput}
R> sigmahat <- estim_sigma(Xsim$X, k = 10, method = "LN")
\end{CodeInput}
\begin{CodeOutput}
R> sigmahat
[1] 0.00079
\end{CodeOutput}
\begin{CodeInput}
R> sigmahat <- estim_sigma(Xsim$X, method = "LN")
\end{CodeInput}
\begin{CodeOutput}
[1] "k =  10"
Warning message:
In estim_sigma(Xsim$X, method = "LN") :
Since you did not specify k, k was estimated using the FactoMineR estim_ncp function
\end{CodeOutput}
Both estimations of $\sigma$ with MAD (0.00811) and with LN (0.0079) are close  since both asymptotic assumptions are plausible: the rank is quite small in comparaison to the size of the data and the strength of the signal is important.
%But of course, results may differ. 

Other options of \code{adashrink} are available:
\begin{CodeInput}
R> adashrink <- function(X, sigma = NA, method = c("GSURE", "QUT", "SURE"), 
method.optim = "BFGS", gamma.seq = seq(1, 5, by=.1), lambda0 = NA, 
center = "TRUE", nbsim = 500)
\end{CodeInput}

If $\sigma^2$ is known, it can be passed in the argument \code{sigma}.
SURE  \eqref{eq:SURE} and GSURE \eqref{eq:GSURE} are optimized using the \code{optim} function in \proglang{R} with its optimization method \code{method.optim = "BFGS"}.
Minimization over $(\lambda, \gamma)$
is performed continously over $\lambda$ and discretly over a  grid for $\gamma$ (defined in \code{gamma.seq}); 
the initial value for $\lambda$ can be given in \code{lambda0} and must be on the log scale (by default the median of the singular values).
The argument \code{center = "TRUE"} is used to center the matrix~$X$. Finally, the argument  \code{nbsim} denotes the number of Monte Carlo runs used when
\code{method = "QUT"} to evaluate the regularization parameter $\lambda^{\rm{QUT}}$ defined as some upper quantile of a distribution sampled by Monte Carlo.

%and is evaluated empirically by a Monte Carlo simulation which generates \code{nbsim} times data under the null hypothesis of no signal (that is,  $\mu = O$).

With \code{adashrink},   soft-thresholding \eqref{eq:soft} can be enforced by setting \code{gamma.seq = 1},
otherwise SURE and GSURE  adapt thresholding to the data with  $\gamma$. For instance we expect $\hat \gamma$ close to 1
when the signal is overwhelmed by noise: % which is the regime of predilection of the soft-thresholding approach:
\begin{CodeInput}
R> Xsim <- LRsim(n = 200, p = 500, k = 100, SNR = 0.5)
R> ada.gsure <- adashrink(Xsim$X, method = "GSURE")
\end{CodeInput}
\begin{CodeOutput}
R> ada.gsure$gamma
[1] 1.1
\end{CodeOutput}

In  \pkg{denoiseR}, the function \code{optishrink} estimates the signal with estimators  \eqref{eq:shaba} and \eqref{eq:verb}:
\begin{CodeInput}
R> optishrink <- function(X, sigma = NA, center = "TRUE", method = c("ASYMPT", "LN"), 
loss = c("Frobenius", "Operator","Nuclear"),  k = NA)
\end{CodeInput}
The  \code{method = "ASYMPT"} corresponds to the asymptotic framework of \eqref{eq:shaba}
%($n$ and $p$ tend to infinity  while the rank $k$ stays fixed).
where the optimal shrinker depends on the loss used,  with \code{loss = "Frobenius"} by default.
The other arguments are the same as before.
%For instance when \code{method = "LN"} and no value for sigma is provided, we use \eqref{eq:sigma_hat} and infer $k$ as well if it is not specified.
The outputs are the same as those of \code{adashrink}.

\subsection{Using denoiseR for non-Gaussian noises}

We recommand \code{ISA} for such a situations, with the following options:
\begin{CodeInput}
ISA <- function (X, sigma = NA, delta = NA, noise = c("Gaussian", "Binomial"), 
transformation = c("None","CA"), svd.cutoff = 0.001, maxiter = 1000, 
threshold = 1e-06, nu = min(nrow(X), ncol(X)), svdmethod = c("svd", "irlba"), 
center = TRUE)
\end{CodeInput}

The CA transformation \eqref{eq:ca} can be obtained by setting the argument \code{transformation = "CA"}; by default it is \code{transformation = "NONE"}.
In the CA case, the noise model is set to  \code{noise = "Binomial"}  by default. 

If the argument \code{delta} is not specified, it is set to $0.5$  by default.
Otherwise, it is estimated for the Binomial noise by repeated learning cross-validation  using the function \code{estim\_delta}:
\begin{CodeInput}
R> estim_delta (X, delta = seq(0.1, 0.9, length.out = 9),  nbsim = 10,  
noise = "Binomial",transformation = c("None", "CA"),  pNA = 0.10, 
maxiter = 1000, threshold = 1e-08))
\end{CodeInput}
This function returns a matrix \code{msep} with the prediction error obtained for $\delta$ varying from 0.1 to 0.9, as well as the value of $\delta$ minimizing the mean squared errors prediction in the object \code{delta}. The argument \code{pNA} indicates the percentage of missing values inserted and predicted  and the argument \code{nbsim}  the number of times this process is repeated. 

\code{ISA} can also be applied with \code{noise = "Gaussian"}  for model \eqref{mod:gauss}. If $\sigma$ is not specified, it is estimated by default with the function \code{estim\_sigma} using the argument \code{method = "MAD"}. 
The option \code{maxiter} corresponds to the maximum number of iterations of \code{ISA} whereas \code{threshold} is for assessing convergence (difference between two successive iterations).
It is possible to specify \code{svdmethod = "irlba"}
to use a fast SVD which is particularly useful when dealing with large matrices. 
In this case, the number of computed singular values can be specified with the argument \code{nu}.

\code{ISA} returns  the estimation in \code{mu.hat}, the number of non-zero singular values in \code{nb.eigen}, the results of the SVD of the estimation in \code{low.rank}
and the number of iterations taken by \code{ISA} in \code{nb.iter}.

%%%%%%%%%%%%%%%%%%%%%%%%%%%%%%%%%%%%%%%%%%
\section[]{Missing values} \label{sec:comp}

We extend and implement here all estimators to the presence of missing data. As an aside, we get new ways to impute data. Both extensions of ATN and ISA are new.
Another side contribution is to extend SURE to missing values. 

\subsection{Iterative imputation} \label{sec:isvd}

Low rank matrix estimation using \eqref{eq:tsvd} or \eqref{eq:soft} has been extended for an incomplete data set \citep{JosseHusson12, Trevor15}
by replacing the least squares criterion by a weighted least squares
\begin{equation} \label{eq:softmiss}
\argmin_\mu  \left\{ \frac{1}{2}\Norm{ W\odot(X- \mu)}_2^2 +  \lambda {\Norm{\mu }_{*}}\right\},
\end{equation}
where $W_{ij}=0$ when $X_{ij}$ is missing and 1 otherwise and $\odot$ stands for the elementwise multiplication.
In the \proglang{R} package \pkg{softImpute} \citep{softimpute},  \citet{Trevor15}
solved equation \eqref{eq:softmiss} using an iterative imputation algorithm. Such an algorithm starts by replacing the missing values by initial values such as the mean of the non-missing entries,
then the estimator is computed on the completed matrix and the predicted values of the missing entries are updated using the values given by the new estimation.
%(observed values are the same and the missing ones are replaced by the fitted values).
The two steps of estimation and imputation are iterated until empirical stabilization of the prediction. 
At the end, since the algorithm denoises the signal but also imputes the missing entries, %on which any method can be applied.
it can be seen as a matrix completion and single imputation method \citep{Schafer97, Little02}.
%Different families of algorithms have been suggested to solve \eqref{eq:softmiss},
%either based on alternating weighted least squares or on iterative imputations. 
%The later class of algorithms took the lead since, in addition to obtaining an estimator for the signal $\mu$,
%they directly provide a completed data set on which any method can be applied.
We follow the same rationale and define iterative imputation algorithms for ATN in Algorithm~\ref{alg:iterATN}.
The \textit{iterative ISA algorithm} can readily be defined. 
% \eqref{eq:sard} and ISA (Algorithm \ref{alg:iter}).
% For given regularization parameters $(\lambda,\gamma)$,  the \textit{iterative ATN algorithm} is defined in Algorithm~\ref{alg:iterATN}.
\begin{algorithm}[h]
\caption{Iterative ATN imputation algorithm for given regularization parameters $(\lambda,\gamma)$.}
\label{alg:iterATN}
\begin{algorithmic}
\STATE Initialization $\ell=0$: substitute missing values with initial values and denote by $X^0$ the initial matrix. %Calculate $M^0$, the matrix of the vector containing the mean of the variables of $X^0$, repeated in each row of $M^0$.
\FOR{$\ell \geq 1$}
  \STATE (a) SVD  of $X^{\ell -1}$ to estimate quantities $d^{\ell}$ and $u^{\ell}$, $v^{\ell}$ and  compute the fitted matrix with ${\hat \mu}_{{(\lambda, \gamma)}_{ij}}^{\ell}  = \sum_{l=1}^{min(n-1,p)} d_l \max\left(1-\frac{\lambda^\gamma}{d_l^\gamma},0\right) u_{il}^{\ell}   v_{jl}^{\ell}$ 
\STATE (b) define the new imputed data as $X^{\ell}= W \odot X + ({\bf1}- W)\odot {\hat \mu}_{{(\lambda, \gamma)}}^{\ell}$, 
where ${\bf 1}$ is a matrix of size $n\times p$ filled with ones. 
\ENDFOR
  \STATE  Stop when $\sum_{ij}({\hat \mu}_{{(\lambda, \gamma)}_{ij}}^{\ell-1} -{\hat \mu}_{{(\lambda, \gamma)}_{ij}}^{\ell} )^2\leq \varepsilon$, with $\varepsilon$ equal to $10^{-6}$ for example.
\end{algorithmic}
\end{algorithm}
% In the same way, we also define an \textit{iterative ISA algorithm} where we alternate between imputation and estimation.
%The procedure being computationally intensive for ISA, we advocate its use  only to carry out the cross-validation technique described in Section~\ref{sec:boot} to estimate $\delta$. 
Iterative ATN and ISA  have not been investigated theoretically yet, but a simulation study in Section \ref{sec:completion} shows that the good empirical properties
observed in the complete case seem to extend to the missing data case.

% we expect good recovery  thanks to to their ability to accurately estimate the signal in the complete case. We checked these claims using a small simulation study detailed Section \ref{sec:completion}.

\subsection{SURE with missing values} 

The \proglang{R} package \pkg{softImpute} \citep{softimpute} does not provide a method to select the regularization parameter $\lambda$ but suggests using cross-validation. 
In \pkg{denoiseR}, %in addition to providing algorithms for other estimators (ATN and ISA),
we suggest a selection of the regularization parameters with missing values based on the Stein unbiased risk estimate. 
In the complete case, we used SURE as an alternative to cross-validation as it is less computationaly costly. SURE is designed to select the parameters $(\lambda, \gamma)$ that minimize the risk of the estimator.
%The bottleneck is to derive the divergence of the estimator.
Since no close form solution is available for the divergence of the estimator when missing values are present, SURE has, as far as we know, never been defined with missing values.
We circumvent the problem by adapting the SURE formula \eqref{eq:SURE} criterion for missing values: the first term is replaced  by $-(np -|{\rm NA}|) \sigma^2$, where $|{\rm NA}|$ the number of missing cells;
the second term is replaced by the RSS on the observed values $\sum_{ij \in obs}  (X_{ij} - \hat \mu_{(\lambda,\gamma)_{ij}}^{\mbox{miss}})^2$,
where $\hat \mu_{(\lambda,\gamma)}^{\mbox{miss}}$ is the estimation and imputation obtained with Algorithm \ref{alg:iterATN};
the third term involving the divergence is calculated using finite differences with
\begin{eqnarray}
\label{eq:divna}
{\rm div^{\mbox{miss}}}(\hat \mu_{(\lambda,\gamma)}^{\mbox{miss}} ) = \sum_{ij \in obs}\frac{ \{ \hat \mu_{(\lambda,\gamma)}^{\mbox{miss}} (X+ \delta {\bf1}_{ij}) \}_{ij} - \{ \hat \mu_{(\lambda,\gamma)}^{\mbox{miss}} (X)\}_{ij}}{\delta},
\end{eqnarray}
where $\delta$ a small variation near machine precision and ${\bf1}_{ij}$ is the matrix of size $n\times p$ with 0 except a 1 at position $\{ij\}$. Consequently,
$\{\hat \mu_{(\lambda,\gamma)}^{\mbox{miss}} (X+ \delta{\bf1_{ij}}) \}_{ij}$ is obtained by  first applying the iterative algorithm \ref{alg:iterATN} to the data matrix $X$ with $\delta$ added to its $\{ij\}$ entry and then by keeping the estimated value for the entry $\{ij\}$.
The resulting SURE formula for missing values is 
\begin{equation} \label{eq:sureatnna}
{\rm SURE^{\mbox{miss}}} = - (np -|NA|) \sigma^2 +
\sum_{ij \in obs}  (X_{ij} - \hat \mu_{(\lambda,\gamma)_{ij}}^{\mbox{miss}})^2+ 2 \sigma^2 {\rm div^{\mbox{miss}}}(\hat \mu_{(\lambda,\gamma)}^{\mbox{miss}}).
\end{equation}
 GSURE  \eqref{eq:GSURE} can readily be extended to missing values to deal with unknown noise variance.
SURE with finite differences was also discussed in \citet{Ramani08} and \citet{Gab14}. The former used it for a variety of estimators and appreciated the ``black-box" aspect of the technique since no knowledge of the functional form of the estimators is required to compute it,
while the latter focused on the gradient of SURE criteria to optimize it with quasi-Newton algorithms.

\begin{figure} 
    \begin{center}
    \includegraphics[width=0.4\textwidth]{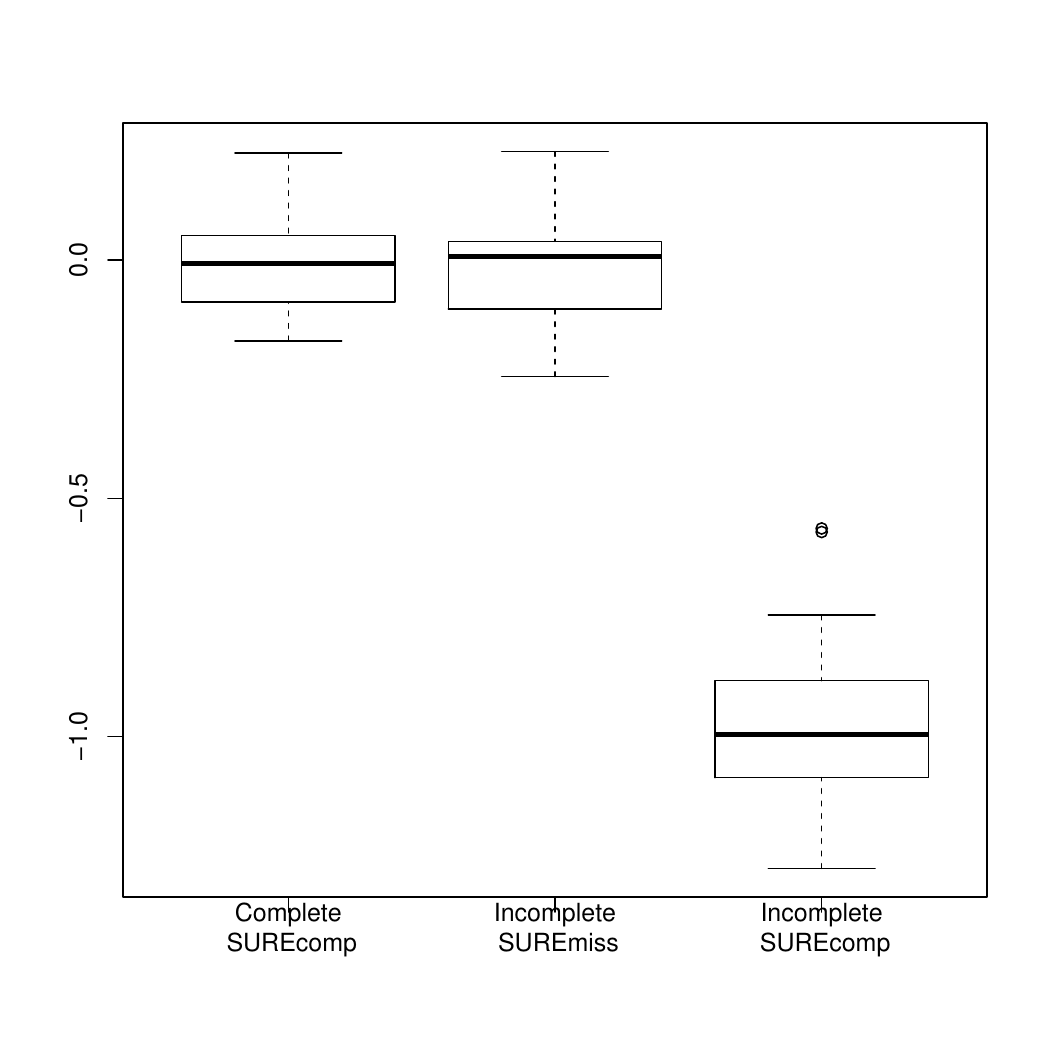}
 \caption{Boxplots of the biais for SURE: on a complete data set (left), with missing data using ${\rm SURE^{\mbox{miss}}}$  (middle),
 with missing data using ${\rm SURE^{\mbox{comp}}}$ on imputed data (right). \label{fig:biais}}
% (MSE$^{\mbox{comp}}$ - SURE$^{\mbox{comp}}$), (MSE$^{\mbox{miss}}$ - SURE$^{\mbox{miss}}$)  and  (MSE$^{\mbox{impute}}$ - SURE$^{\mbox{comp}})$
    \end{center} 
  \end{figure}

To assess the validity of our proposal, we check that \eqref{eq:sureatnna} is an unbiased estimate of the risk on a simulation for $\gamma=1$ (i.e., the soft-thresholding estimator) 
and known noise variance $\sigma^2$.
We  consider a complete data generated according to \eqref{mod:gauss} with 50 rows, 30 colums, rank $k=5$ and SNR of 0.5. 
Then, for a given value of $\lambda$, we compute the \textit{complete} estimator $\hat \mu_{\lambda}$ \eqref{eq:soft}, its risk  MSE$^{\mbox{comp}}$ = MSE $(\hat \mu_{\lambda}, \mu)$ and the estimator of the risk with  SURE$^{\mbox{comp}}$ \eqref{eq:SURE}.
Next, we create 20\% missing entries completely at random and impute the missing values with $\hat \mu_{\lambda}^{\mbox{miss}}$ using the \textit{iterative ATN algorithm} described in Section \ref{sec:isvd}.
To estimate the corresponding risk, we use either SURE$^{\mbox{miss}}$ \eqref{eq:sureatnna} to take imputation into account (working only on the observed values)
or SURE$^{\mbox{comp}}$ which treats the imputed data as if they were real observations. 
We repeat this procedure one hundred times and represent the distribution of the difference between the true loss and SURE in Figure~\ref{fig:biais}.
As expected, SURE is an unbiased estimate of the risk in the complete case (left),
and we see that our risk estimate \eqref{eq:sureatnna}  corrects well the biasedness  of a simple SURE formula employed as if  the imputed data were observations (right).
% as well as in the incomplete case \eqref{eq:sureatnna}. The last boxplot represents the case where the data were imputed with the ATN estimator and the risk estimated with  SURE$^{\mbox{comp}}$. It highlights that it is clearly not suitable to consider the imputed data set as a ``true" one and to forget about the missing values.
Comparable results were obtained with other matrix sizes and values of $\lambda$ and $\gamma$.

Despite the theoritical soundness of this approach, we should remark that it is computationaly heavy so that the computional advantage over cross-validation advocated in the complete case is no longer so clear. We investigate this point in the next section.
%In addition, finite-differences could be unstable. 

\subsection{Application to matrix completion} \label{sec:completion}

A small simulation study shows the potential of the \textit{iterative ATN algorithm} to impute data.
We use data sets that differ in terms of number of observations $n$, number of variables $p$ and relationships between variables.
More precisely, the first two data sets are generated from model \eqref{mod:gauss} with rank $k$, respectively equal to 3 and 2, whereas the Parkinson data set \citep{Buhlmann12} is known for its nonlinear relationships between variables.
Next, we insert 20\% of missing values according to the two mechanisms missing completely at random (MCAR) and missing at random (MAR) \citep{Little02}. 
For the former, we simply insert missing values uniformally whereas for the latter, we put missing values in one variable say $X_1$ when the values of another variable say $X_2$ is greater than the upper quartile of $X_2$.
We use this approach for different variables and different quartile until reaching a desired percentage of missing entries. The detail is available in the associated code provided as supplementary material.  

We impute the data with the following methods:
\begin{itemize}
\item ATN  based on GSURE.
\item soft thresholding \eqref{eq:softmiss}  by \citet[package \pkg{softImpute}]{softimpute}.
\item random forests (RF)  by \citet[package \pkg{missForest}]{Buhlmann12}.
\item based on Gaussian distribution  by \citet[package \pkg{mice}]{mice, Buur12}. Imputation is obtained using iterative imputation with multiple regressions. 
\item a baseline imputation with the mean of the variables on the observed values.
\end{itemize}
We then compute the squared error of prediction,  and repeat the process 50 times. Table \ref{tablepar} reports the mean squared errors of prediction. 
% We  have selected both regularized parameters for ATN with GSURE to be as close as possible to real data analysis with no knowledge for the noise variance.
Since, no method is available to select the regularization parameter $\lambda$ in \pkg{softImpute},
we generate a grid for $\lambda$ and use its oracle value, by minimizing the mean squared errors prediction. 
We generate $m=20$ multiple imputed datasets in \pkg{mice} and take the average for single imputation. The code  to reproduce the simulations is available as supplementary materials.

\begin{table}[h!]
%\scriptsize
\centerline{
\begin{tabular}{|r|r|r|r|r||c|c|c|c|c|}
\hline
Data&$n$&$p$& $k$ & scale &ATN & softImp & RF & mice & Mean \\ \hline
Simulation 1 &30&50 & 3 & (1.e-5)   &\textbf{60}/\textbf{62}  &  67/67& 62/65& NA/NA & 68/71  \\ 
Simulation 2 & 41 & 10 & 2 & (1.e-3) &\textbf{9/10} & 11/22& 33/191 &  10/13 & 223/471\\
Parkinson& 195 & 22 & & (1) & 111/95 & 219/85 & \textbf{79}/\textbf{47}& NA/NA & 111/95\\\hline
\end{tabular}
}
\protect\caption{Mean squared errors of prediction for MCAR (left)/MAR (right). Smallest values row-wise are in {\bf bold}. \label{tablepar}}
\normalsize
\end{table}

\normalsize
The small simulation results show that the proposed methods have strengths and weaknesses.
The first two methods, namely ATN and softImpute, have errors in the same range of magnitude but with a slight advantage for ATN.
This behavior is expected since ATN often improves on softImpute for complete case data \citep{sardy15, JosseWager}.
Both methods perform best under model~\eqref{mod:gauss}  but struggle with non-linear relationships.
Indeed, the imputation for both methods is based on low rank assumption and linear relationship between variables.
 Conversely,  imputation with random forests can cope with non-linearity in the Parkinson data.
 However, despite its recent popularity \citep{Buhlmann12, Doove14, Shah14},  imputation with random forests breaks down for small sample size and MAR cases,
 as already observed in \citet{Audigier13}, because extrapolation and prediction outside of range of data seems difficult with random forests. 
The imputation based on regression (mice)  breaks down  when $n$ smaller than $p$ and for Parkinson data. 
These results are not surprising since the properties of an imputation method depend on the inherent characteristics of the method.  Regression based imputation methods also  encounter difficulties when the variables are highly correlated. Some imputation based on ridge regression have been suggested in \citet{mice} to tackle this issue.

Since the structure of the data  is not known in advance, one could use cross-validation and select the method which best predicts the removed entries.
Nevertheless, we argue that many data sets have a good low rank approximation  \citep{Udell17}, so that imputation based on SVD  often proves accurate.

However, ATN requires to select regularization parameters with GSURE which is computationaly costly as it uses finite differences. To investigate the gain of using GSURE instead of cross-validation, we perform additional simulations under model  \eqref{mod:gauss} varying size of the data, rank $k$ and percentage of missing entries. 
We report a subset of the results in Table \ref{tabletime}.

%\begin{table}[h!]
%\label{tabletime}
%\scriptsize
%\centerline{
%\begin{tabular}{|r|r|r|r|r||c|c|c|c|c|}
%\hline
%Data&$n$&$p$& $k$ & scale &ATN & softImp & RF & mice & Mean \\ \hline 
%Simu 1 &30&50 & 3 & (1.e-5)   &1.39 $\times 10^5$/1.37 $\times 10^5$  &  5.43$\times %10^2$/5.47$\times 10^2$& 1.28$\times 10^3$/9.34$\times 10^2$& NA/NA & 1.31/1.44 \\ 
%Simu 2 & 41 & 10 & 2 & (1.e-3) &/ & /& / &  / & /\\
%Parkins& 195 & 22 & & (1) & / & / & /& NA/NA & /\\\hline
%\end{tabular}
%}
%\protect\caption{Time in milli-second of one simulation for MCAR (left) and MAR (right). %\label{tabletime}}
%\end{table}

\begin{table}[h!]
%\scriptsize
\centerline{
\begin{tabular}{|r|r|r|r|r||r|r|r|r|}
\hline
\multicolumn{5} {|c||} {10\% missing} & \multicolumn{4} {c|} {20\% missing} \\
\hline
 $(n,p)$&$(50,5)$&$(50,5)$&$(100,5)$&$(100,5)$  &$(50,5)$&$(50,5)$& $(100,5)$  &$(100,5)$ 
 \\\hline
     &MSE& time  &MSE& time  &MSE& time &MSE& time\\
RF     &  0.074  &0.168 & 0.0719   &  1.451 &0.159  & 0.303&0.156  & 0.508\\
%IPCA      &0.057   &0.008 & 0.0586   & 0.050  & 0.122  & 0.025&0.141  & 0.021\\
%mice      &0.054   &0.004  & 0.0566  & 0.0027 & 0.121  & 0.007&0.128  & 0.006\\
Mean     & 0.141  & 0.001  & 0.1264     & 0.001 & 0.262  & 0.000&0.259  & 0.001\\
ATN       &0.052 &132.572 &0.0567&8371.077 & 0.111 &248.359& 0.126& 462.570\\
ATN CV        &0.053   &5.314 & 0.0571   &  63.076 & 0.113 & 11.009&0.127 &  9.986\\
%softImpCV    &0.070&   0.612 &  0.0689  &  3.225&  0.142 &  1.134&0.157   &0.975\\
\hline
\end{tabular}
}
\protect\caption{MSE and average time in second over 50 simulations for data of size $(n,p)$ generated under  \eqref{mod:gauss} with rank 2 and for 10\%  and 20\% of missing values. Comparison of imputation with random forest, ATN with GSURE and ATN with cross-validation (CV).\label{tabletime}}
\end{table}

Although ATN with GSURE still gives the smallest MSEs, the method is extremelly costly and takes more time than ATN with cross-validation (we implemented a 10-fold cross validation) which gives MSEs in the same order of magnitude.
Imputation with random forests is also reported: it is known to be slow, but still faster than ATN. However, as observed in Table \ref{tablepar}, ATN still provides the best prediction of missing values and improves on the competitors (Random forests, SoftImpute, mice). 

Table \ref{tabletimediff} shows that, in very difficult settings, ATN with GSURE may encounter more difficulties than ATN with CV. \cite{Ef2004} shows that both approaches have advantages and drawbacks but that  SURE may offer better accurancy when the model is correct. 
\begin{table}[h!]
%\scriptsize
\centerline{
\begin{tabular}{|r|r|r|}
\hline
 $(n,p)$&$(30,50)$ &(30,50)
 \\\hline
     &MSE& time \\
RF    &  0.110  &  1.451 \\
%IPCA      &0.060   & 0.050  \\
%MICE      &0.054   &0.004  & 0.0566  & 0.0027 &NA  &    NA & 0.121  & 0.007&0.128  & 0.006\\
Mean     &  0.135   & 0.001 \\
ATN       &0.127 &8371.077 \\
ATN CV      & 0.060 &  63.076 \\\hline
%CVsoft    &0.070&   0.612 &  0.0689&   0.6316  &0.071  &  3.225&  0.142 &  1.134&0.157   &0.975\\
\end{tabular}
}
\protect\caption{MSE and average time in second over 50 simulations for data of size $(30,50)$ generated under  \eqref{mod:gauss} with rank 10 and 10\% of missing values. \label{tabletimediff}}
\end{table}

%Finally \pkg{mice} has difficulties either due to the strong linear or non-linear relationships. %Indeed, for the former, the regression are unstable.

%Small sample size and MAR cases can not be well handled.

%First, we note that results based on the imputation with the Gaussian distribution are very poor. For the first 2 data, this can be explained %because the variables are highly correlated and thus the regressions are unstable while it is due to the non-linear relationship between %variables for the last one. ATN provides better prediction of the missing values in MCAR and MAR cases and improves on SoftImpute. 

%This is particularly striking for MAR missing values where random forests never succeed due to their inability to ``extrapolate". 
%However, it is the only method which can directly cope with the highly non-linear Parkinson data at least for the MCAR setting. 

\subsection{Implementation}

To perform the \textit{iterative ATN algorithm} on an incomplete data, we use the \code{imputeada}  function  with the following options:
\begin{CodeInput}
R> imputeada(X, lambda = NA, gamma = NA, sigma = NA, method = c("GSURE", "SURE"), 
gamma.seq = seq(1, 5, by=.1), method.optim = "BFGS", center = "TRUE", 
scale = "FALSE", threshold = 1e-8, nb.init = 1, maxiter = 1000, lambda0 = NA)
\end{CodeInput}

If the argument \code{lambda} and  \code{gamma} are not specified, they are estimated with \code{method = "GSURE"} or \code{method = "SURE"}.
Contrarily to the complete case, the argument \code{method = "QUT"}  is not available. Moreover one must specify the variance of the noise when using the argument \code{method = "SURE"} \eqref{eq:sureatnna}
because estimation of $\sigma$ with missing values is not available yet. 
The outputs are the same as for complete data. %, in particular the estimation obtained with missing values is available in the object \code{muhat}.
In addition, the function outputs a completed data matrix in  \code{completeObs}. 
%Even if it is out of the scope of the present paper to focus on this topic, we ran a

\section{Exploratoration and visualization using denoiseR}\label{sec:data}

\subsection{Unsupervised clustering on a tumors data}

%\fbox{Why talk about cluster?} 

\begin{figure}
    \begin{center}
    \includegraphics[width=0.49 \textwidth]{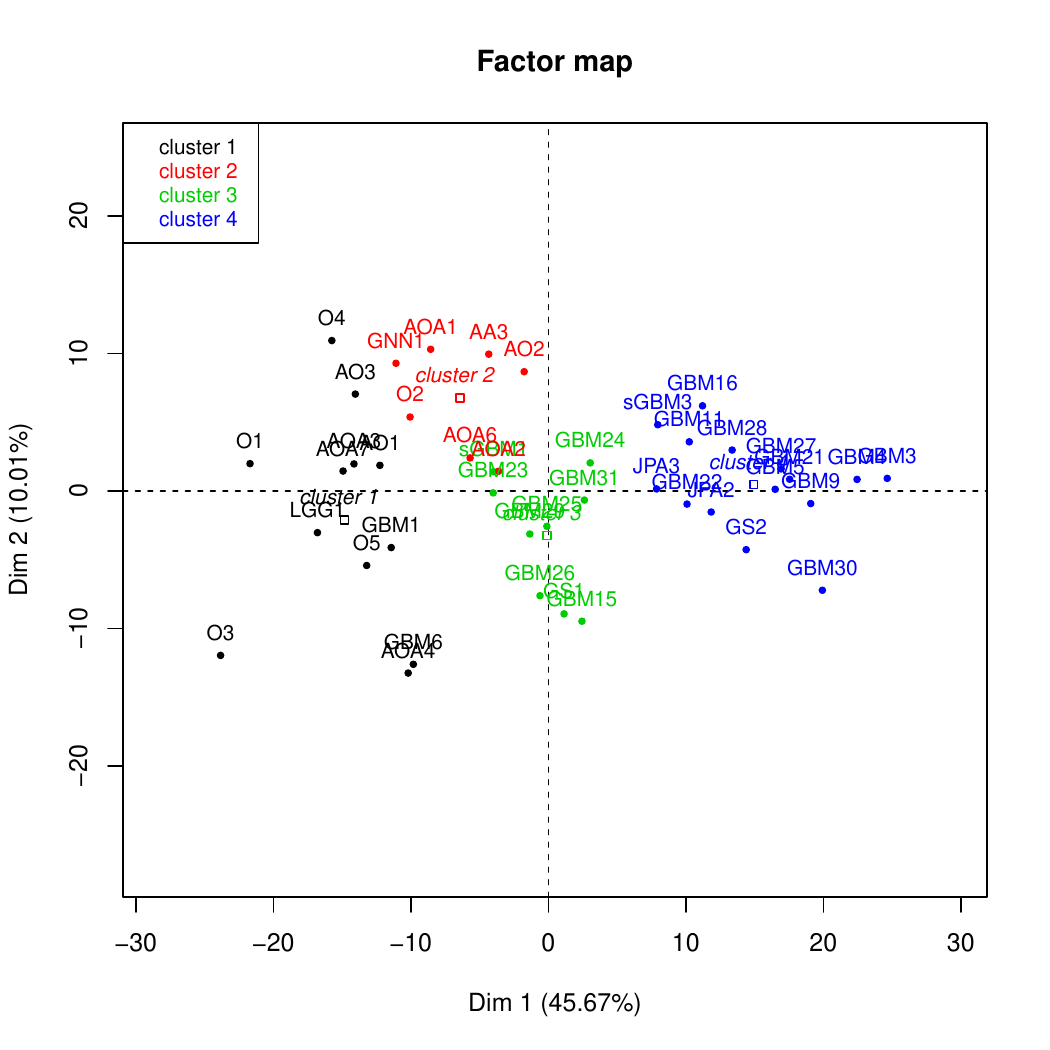}  \includegraphics[width=0.49 \textwidth]{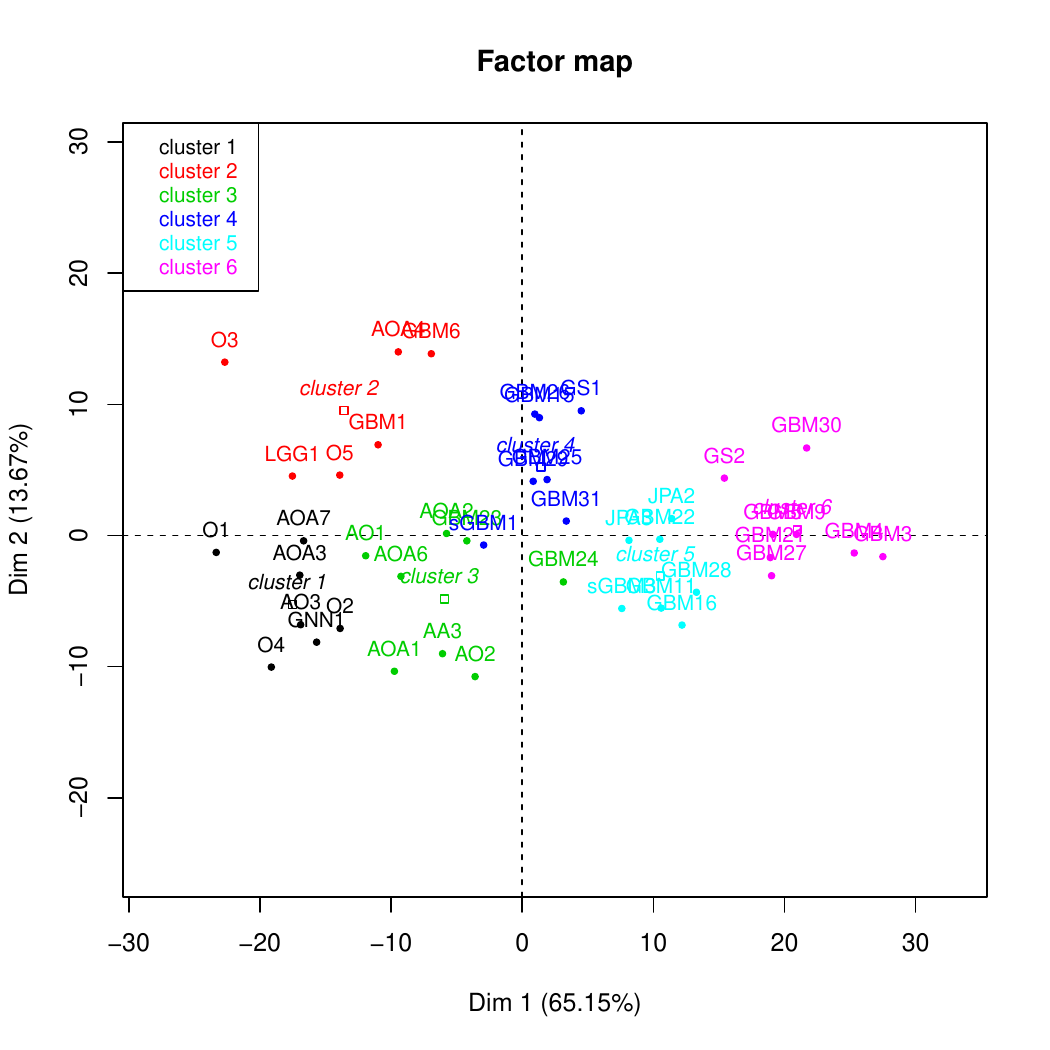}
 \caption{Clustering of the tumors data denoised by truncated SVD (left) and  ATN (right). \label{fig:cluster}}
    \end{center} 
\end{figure}

Unsupervised clustering is often applied on denoised data to better identify the clusters  \citep{Paghuss10}. We illustrate the difference in clustering results after denoising either with truncated SVD or ATN
 on tumors data. % how to complement low rank matrix estimation with unsupervised clustering.
%In practice, we can use \pkg{denoiseR} before applying an unsupervised clustering algorithm. 
% It is common to perform $k$-means algorithms or hierarchical trees on the estimator given by the truncated SVD \eqref{eq:tsvd} \citep{Paghuss10}.
%, or said differently on the first $k$ principal components of principal component analysis (PCA).
% The rationale is to first denoise the data to get better clustering.
%However, if we believe in model \eqref{mod:gauss}, it seems more appropriate to apply the clustering on the estimators described in Section \ref{sec:lrs} rather than on \eqref{eq:tsvd} since the signal is better estimated with the former. 
The data consist of 43 brain tumors of four different types defined by the standard world health organization (WHO) classification (O, oligodendrogliomas; A, astrocytomas; OA, mixed oligo-astrocytomas and GBM, glioblastomas) and 356 continuous variables corresponding to the expression data.  
We start with the classical approach which consists in performing a clustering method on the denoised data estimated with truncated SVD \eqref{eq:tsvd} by means of
 the \pkg{FactoMineR} package which implements principal components methods, unsupervised clustering methods and visualization tools. 
First, we estimate the rank $k$ by cross-validation  \citep{Josse11b} with the function \code{estim\_ncp}. Then, we perform the truncated SVD at $k$ using the \code{PCA} function. Finally, we use a hierarchical clustering on the matrix $\hat \mu_k$ with the \code{HCPC} function. The number of clusters is automatically determined based on the increase of between-clusters variance \citep{Paghuss10}(page 185).
More details on these functions can be found in  \citet{FactoJSS}.  Figure~\ref{fig:cluster} (left) represents the scores of the observations on the first two dimensions of variabilities (the  matrix UD). The points are then coloured with respect to their clusters. 

\begin{CodeInput}
R> data(tumors)
R> nb.ncp <- estim_ncp(tumors[ , -ncol(tumors)])
R> res.pca <- PCA(tumors, ncp = nb.ncp$ncp, quali.sup = ncol(tumors))
R> res.hcpcpca <- HCPC(res.pca, graph = F, consol = FALSE)
R> plot.HCPC(res.hcpcpca, choice = "map", draw.tree = "FALSE")
\end{CodeInput}
Then, we estimate the signal with \code{adashrink} and perform the same clustering algorithm:
\begin{CodeInput}
R> res.ada <- adashrink(tumors[ , -ncol(tumors)], method = "SURE")
R> res.hcpcada <- HCPC(as.data.frame(res.ada$mu.hat), graph = F, consol = FALSE)
R> plot.HCPC(res.hcpcada, choice = "map", draw.tree = "FALSE") 
\end{CodeInput}

We use the \code{SURE} option instead of \code{GSURE} since the latter gave 42 non-zero singular values (the maximum number) which is not very likely.

Figure~\ref{fig:cluster} (right) shows the clustering of this second approach. 
%The impact of regularisation on the graphical representations is not obvious when looking only at the representation of the scores.
The same clustering algorithm applied to the denoised data shows either four clusters with truncated SVD or six clusters with ATN.
% the effect of regularisation becomes ovious when looking at the clustering results. Indeed,
%Since ATN modifies the distances between observations, the ATN clustering will differ from the truncated SVD one.
Results may thus conduct to a different interpretation whether using one option or the other.
The truth is unknown here, but one can expect a better clustering with data denoised by ATN since simulations point to better estimation.

\subsection{Regularized correspondence analysis on a Presidents data} \label{sec:presi}

Let us consider the \code{Presidents} data set, a contingency table cross-tabulating  $13$ US presidents (from 1940 to 2009) with $836$ words used during their inaugural addresses. 
All the texts from the speeches of George Washington in 1789 can be obtained via the public websites \url{http://www.presidency.ucsb.edu} and  \url{http://www.usa-presidents.info/union/}. 
The texts were pre-processessed (lemmatized, keeping words that appear less than 50 times...) to get the $13 \times 836$ contingency table.
The \code{Presidents} data come from  the \proglang{DtmVic} software \citep{dtm} specialized in the analysis of corpus data.

To perform a regularized CA with ISA, we first estimate the amount of bootstrap noise using the function \code{estim_delta} and then we use  the function \code{ISA} with the CA transformation \eqref{eq:ca}. 

\begin{CodeInput}
data(Presidents)
delt <- estim_delta(Presidents, transformation = "CA")
isa.ca <- ISA(Presidents, delta = delt$delta, transformation = "CA")
\end{CodeInput}

The results can be visualized using the \code{CA} and \code{plot.CA} functions in the \pkg{FactoMineR} package:
\begin{CodeInput}
R> rownames(isa.ca$mu.hat) <- rownames(Presidents)
R> colnames(isa.ca$mu.hat) <- colnames(Presidents)
R> res.isa.ca <- CA(as.data.frame(isa.ca$mu.hat), graph = FALSE)
R> plot(res.isa.ca, title = "Regularized CA", cex = 0.7, 
     selectRow = "contrib 40", invisible = "col")
R> plot(res.isa.ca, title = "Regularized CA", cex = 0.8, invisible = "row")
\end{CodeInput}
The argument  \code{invisible} is set respectively to \code{col} or to \code{row} to get two graphs, one for the Presidents and one for the words. The argument \code{selectRow} specifies that only the 40 words which contributes the most to the creation of the dimension of variability are represented. It allows to get graphical representation which are not overloaded. 

 \begin{figure}
    \begin{center}
    \includegraphics[width=0.47 \textwidth, height = 0.3 \textheight]{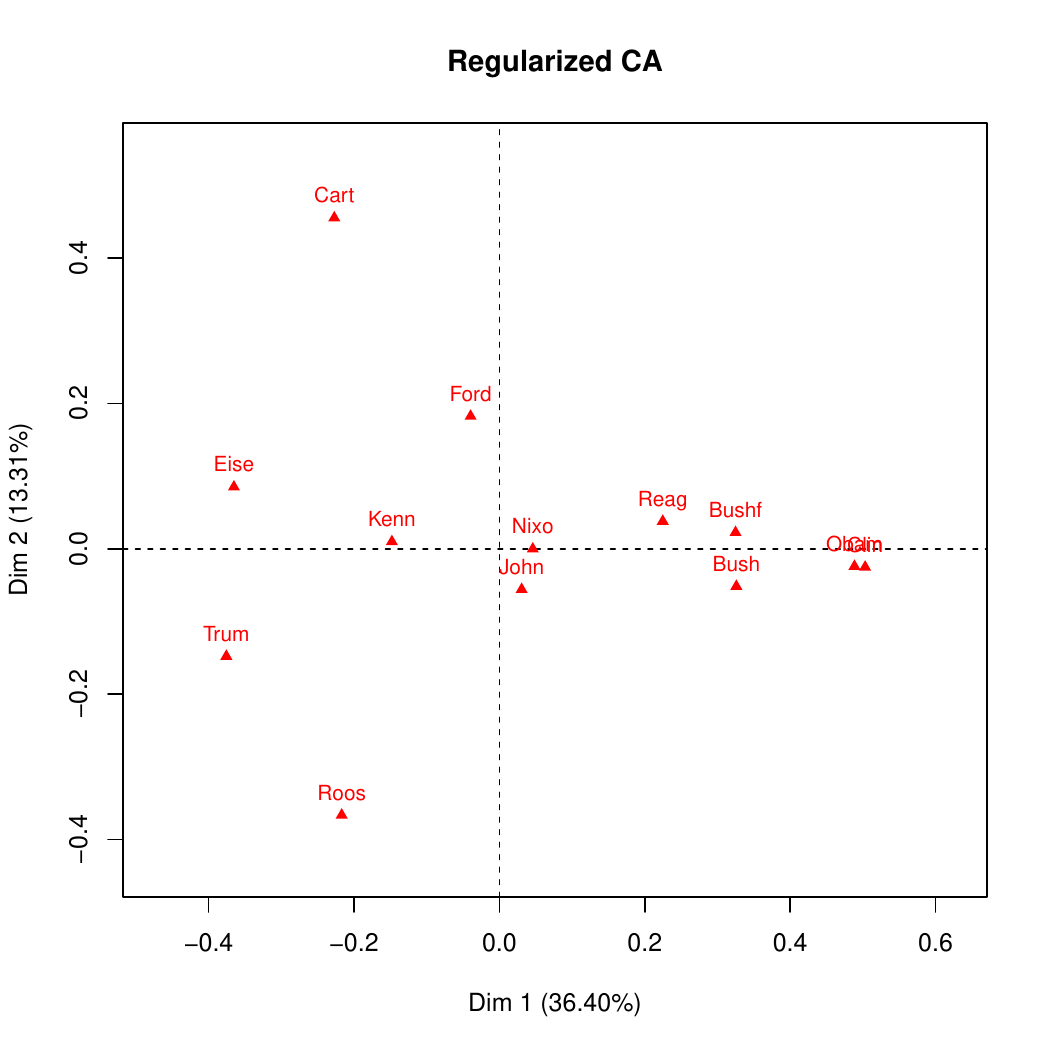}     \includegraphics[width=0.47 \textwidth, height = 0.3 \textheight]{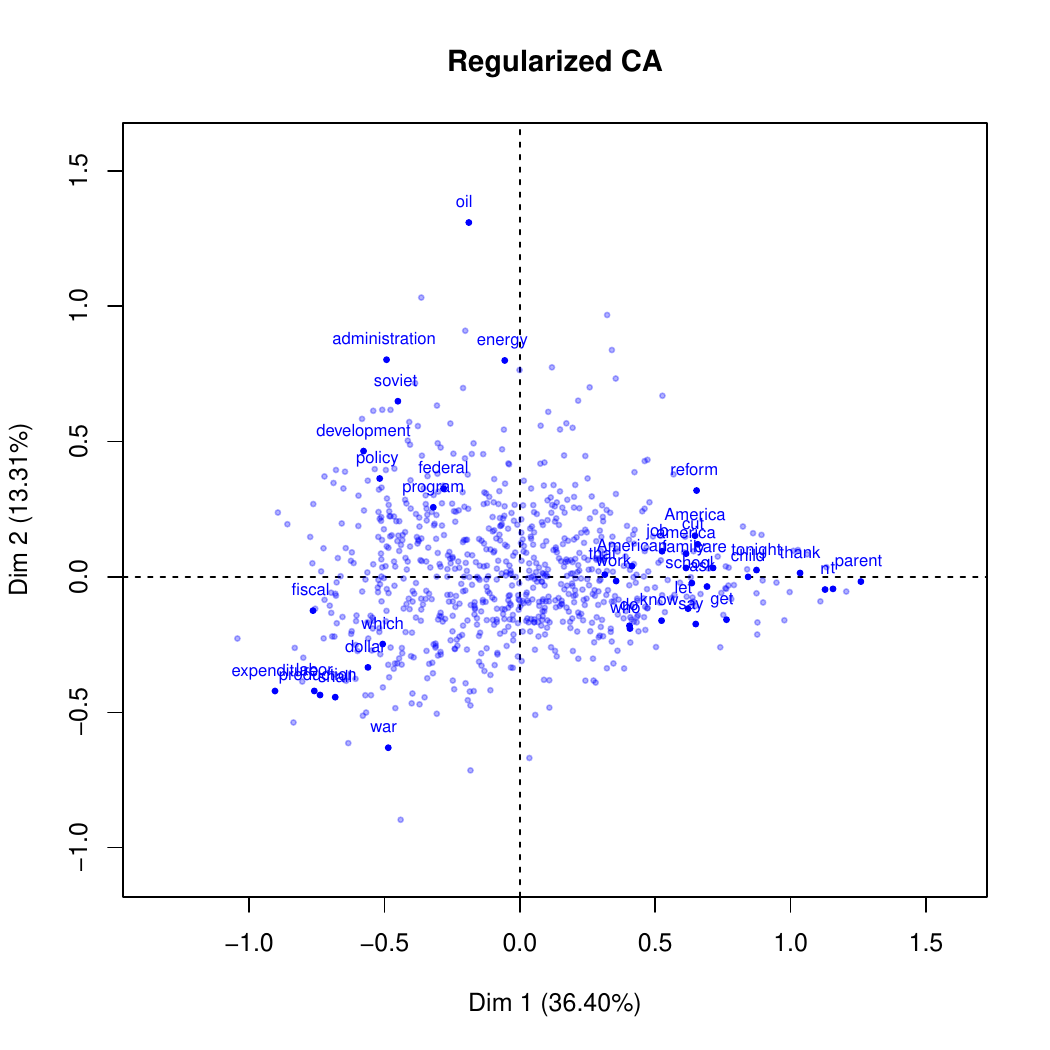}  
 \caption{US Presidents from 1940 to 2009 - Words used during their inaugural speeches.  \label{fig:presi}}
    \end{center} 
  \end{figure}
Figure \ref{fig:presi} represents the presidents and the words on the  first two CA dimensions. Briefly, rules of interpretation are the following: two presidents are close if they have the same profile of words (they over-employed or under-employed the same words); two words are close if there are associated in the same way to the presidents; a president is on the side of the words that he over-employed in comparison to the other presidents.
Thus, we see that at the time of Carter, the words ``soviet", ``administration", ``policies" were often used whereas recently ``family", ``parents", ``child" are often mentioned.
As for the tumors data in Section \ref{sec:finite}, it is also possible to complement the analysis by applying a clustering on the denoised data.

%To do: Add the pictures of the presidents.??

\subsection{Imputation of the impact factor data}

We illustrate the capability of \pkg{denoiseR} to impute on the journal impact factors data from  \url{journalmetrics.com}.
We use a subset of 443 journals of the same sections than Journal of Statistical Software (``Computer Science :: Software", ``Decision Sciences :: Statistics, Probability and Uncertainty" and ``Mathematics :: Statistics and Probability"). This data has 45 columns which correspond to three metrics recorded each year from 1999 to 2013: IPP - impact per publication (it is closed to the ISI impact factor but for three rather than two years), SNIP - source normalized impact per paper (tries to weight by the number of citations per subject field to adjust for different citation cultures) and the
SJR - SCImago journal rank (tries to capture average prestige per publication). 
This data contains 31\% of missing values. We impute it with the \textit{ATN iterative algorithm} selecting the regularization parameters with GSURE on subsamples of data.  

\begin{CodeInput}
R> data(impactfactor)
R> impactfactor[17:24,1:5]
\end{CodeInput}
\begin{CodeOutput}
                              SNIP_1999 IPP_1999 SJR_1999 SNIP_2000 IPP_2000
Annals of Applied Probability      1.519    0.933    1.821     1.336    0.968 
Annals of Applied Statistics          NA       NA       NA        NA       NA
Annals of Probability              1.506    0.920    2.497     1.795    1.132
Annals of Software Engineering     0.529    0.176    0.203     0.690    0.299
Applied Soft Computing Journal        NA       NA       NA        NA       NA   
\end{CodeOutput}
\begin{CodeInput}
R> ada.gsureNA <- imputeada(impactfactor, lambda = 4.5, gamma = 1.9)
R> summary(ada.gsureNA$completeObs)
\end{CodeInput}

After imputation, any analysis can be performed on the completed data.
We apply multiple factorial analysis (MFA) \citep{Escofier98, Pages14} of the \pkg{FactoMineR} package.
MFA can be presented as  a counterpart of PCA for data with groups of variables and it allows to visualize the proximities between journals, the correlation between metrics as well as some trajectories of the journals through the years (relationships between the groups of variables). The core of MFA is a weighted PCA and it provides the same graphical representation than PCA, respectively one plot for the scores and the other for the loadings but in addition provide plots that takes into account the structure of groups of variables.
More details about this method can be found in \cite{FactoJSS} and \citet{Pages14}. 
We use the function \code{MFA} which takes as input the completed data set by iterative ATN. Then, the argument \code{group} specifies that there are 15 groups (years) of respectively 3 columns (metrics) and the argument \code{type} specificies that all the variables of all the 15 groups are scaled ("s") as often  in PCA. The function \code{plot.MFA} allows to represent the plots of the scores of the journal on the 2 first dimensions with \code{ choix = "ind"} and the correlation circle for the variables with \code{ choix = "var"}. The options \code{partial = "all"} with \code{select} set to the name of a specific journal allow to represent the trajectory of a journal through the year. Other lines of codes are just designed to enhance the readibility of the graphics. 

\begin{CodeInput}
R> year = NULL; for (i in 1: 15) year = c(year, seq(i,45,15)) 
R> res.mfa  <- MFA(ada.gsureNA$completeObs, group=rep(3,15),  type=rep("s",15), 
name.group=paste("year", 1999:2013,sep="_"),graph=F)
R> plot(res.mfa, choix = "ind", select = "contrib 15", habillage = "group", cex = 0.7)
R> points(res.mfa$ind$coord[c("Journal of Statistical Software", "Journal of the 
American Statistical Association", "Annals of Statistics"), 1:2],col=2, cex=0.6)
R> text(res.mfa$ind$coord[c("Journal of Statistical Software"), 1], 
res.mfa$ind$coord[c("Journal of Statistical Software"), 2],cex=1, 
labels=c("Journal of Statistical Software"),pos=3, col=2)
R> plot.MFA(res.mfa, choix="var", autoLab="yes", cex=0.5, shadow=TRUE)
R> plot(res.mfa, select="Journal of Statistical Software", partial="all", 
habillage="group",unselect=0.9, chrono="TRUE", xlim=c(-10,20), ylim=c(-10,20))
R> plot(res.mfa, select="IEEE/ACM Transactions on Networking", partial="all",
 habillage="group", unselect=0.9,chrono=TRUE)
\end{CodeInput}

\begin{figure} 
    \includegraphics[width=0.5 \textwidth, height = 0.3 \textheight]{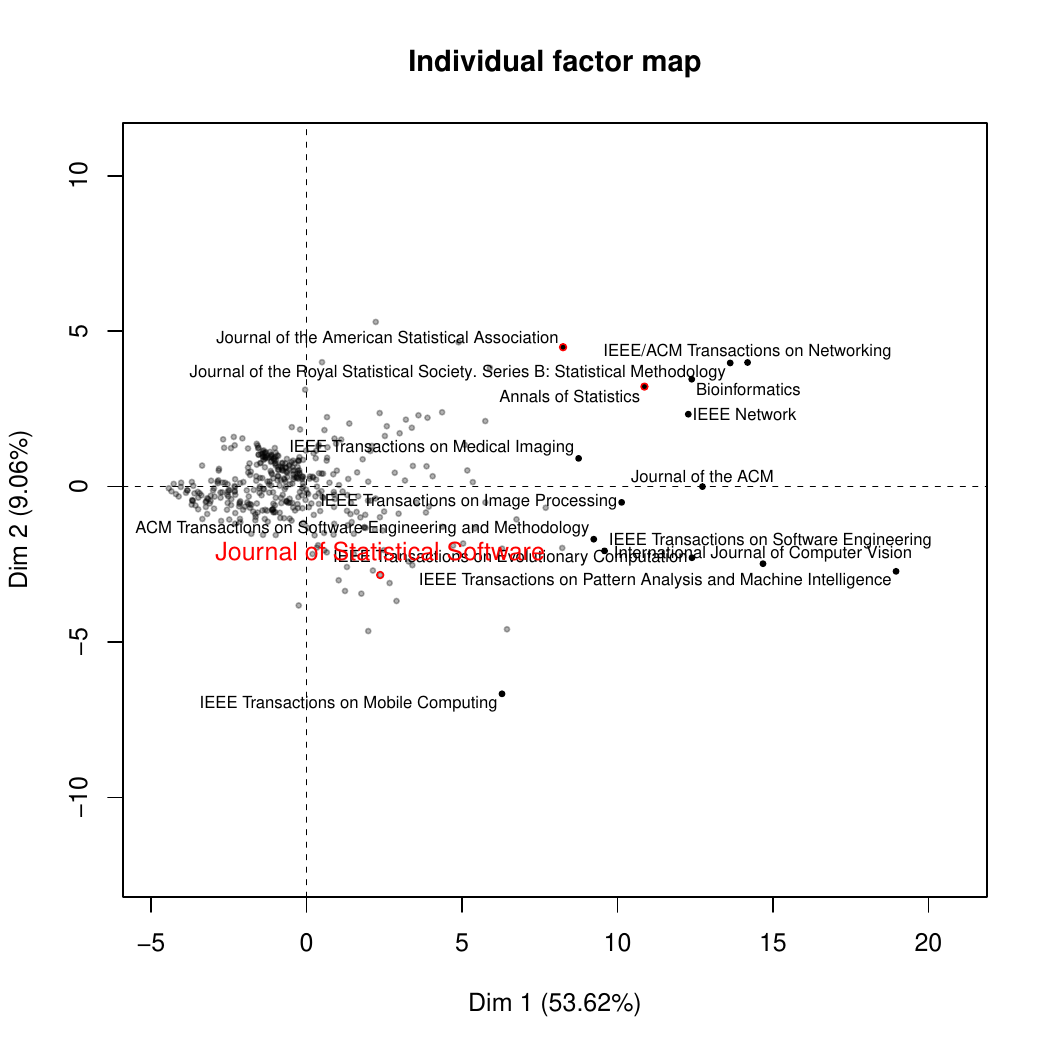}    \includegraphics[width=0.45 \textwidth , height = 0.27 \textheight]{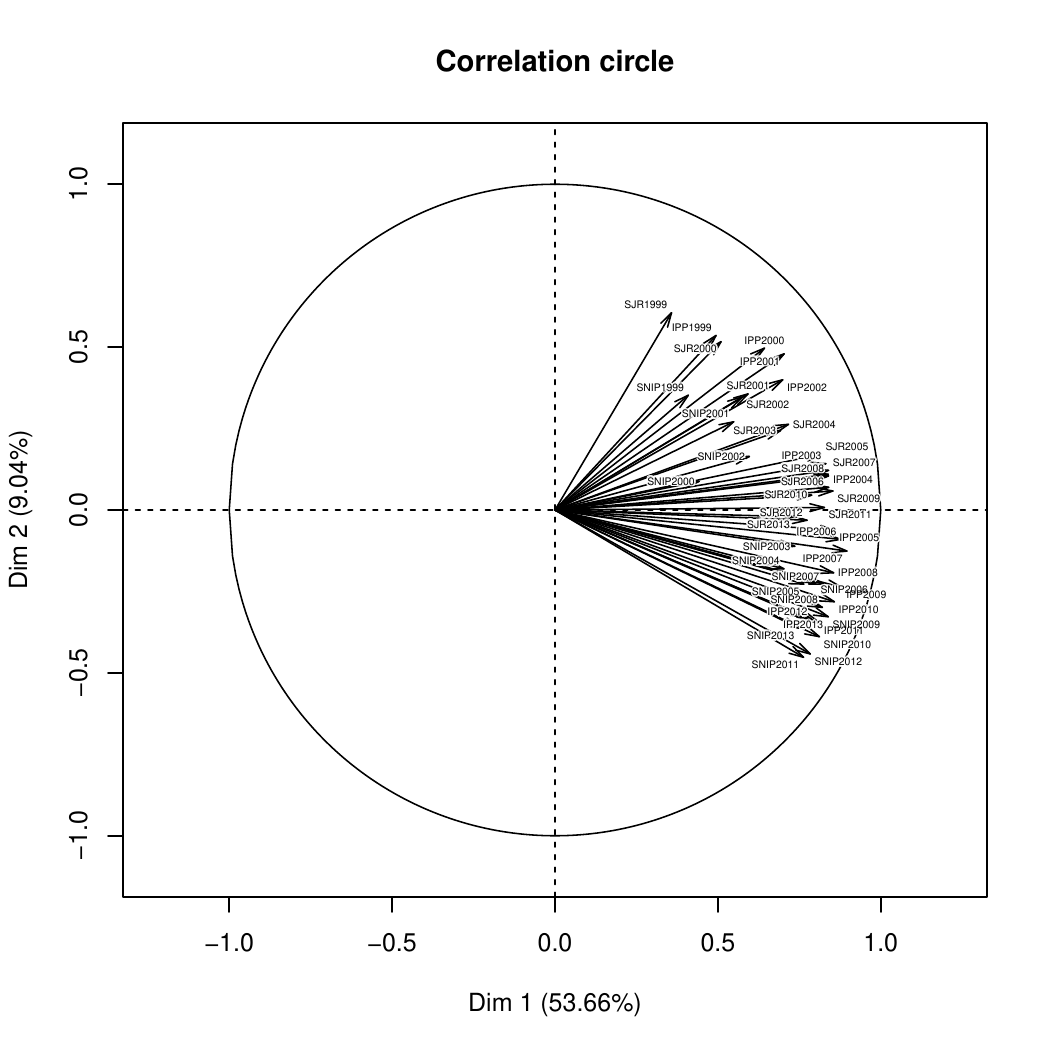} 
    \includegraphics[width=0.45 \textwidth , height = 0.3 \textheight]{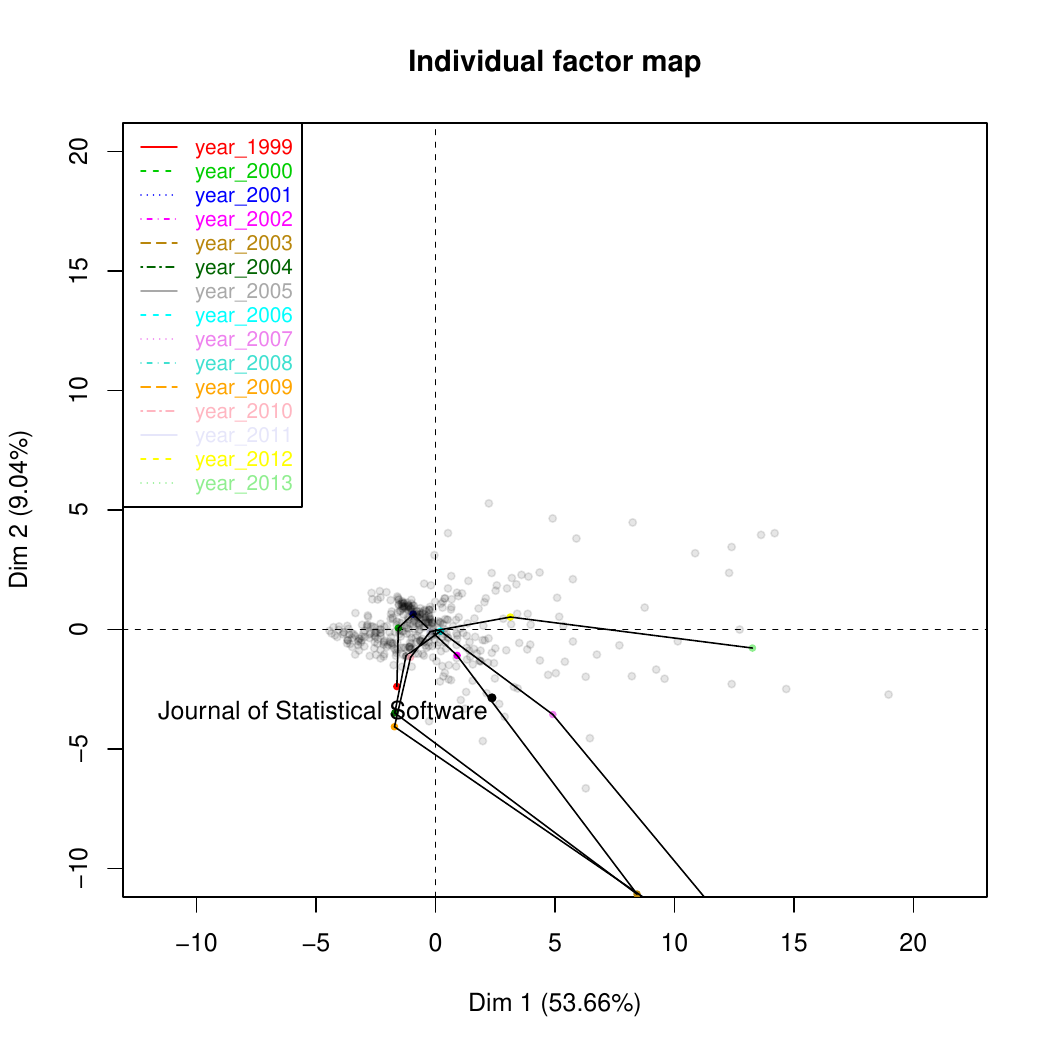}  \includegraphics[width=0.43 \textwidth , height = 0.27 \textheight]{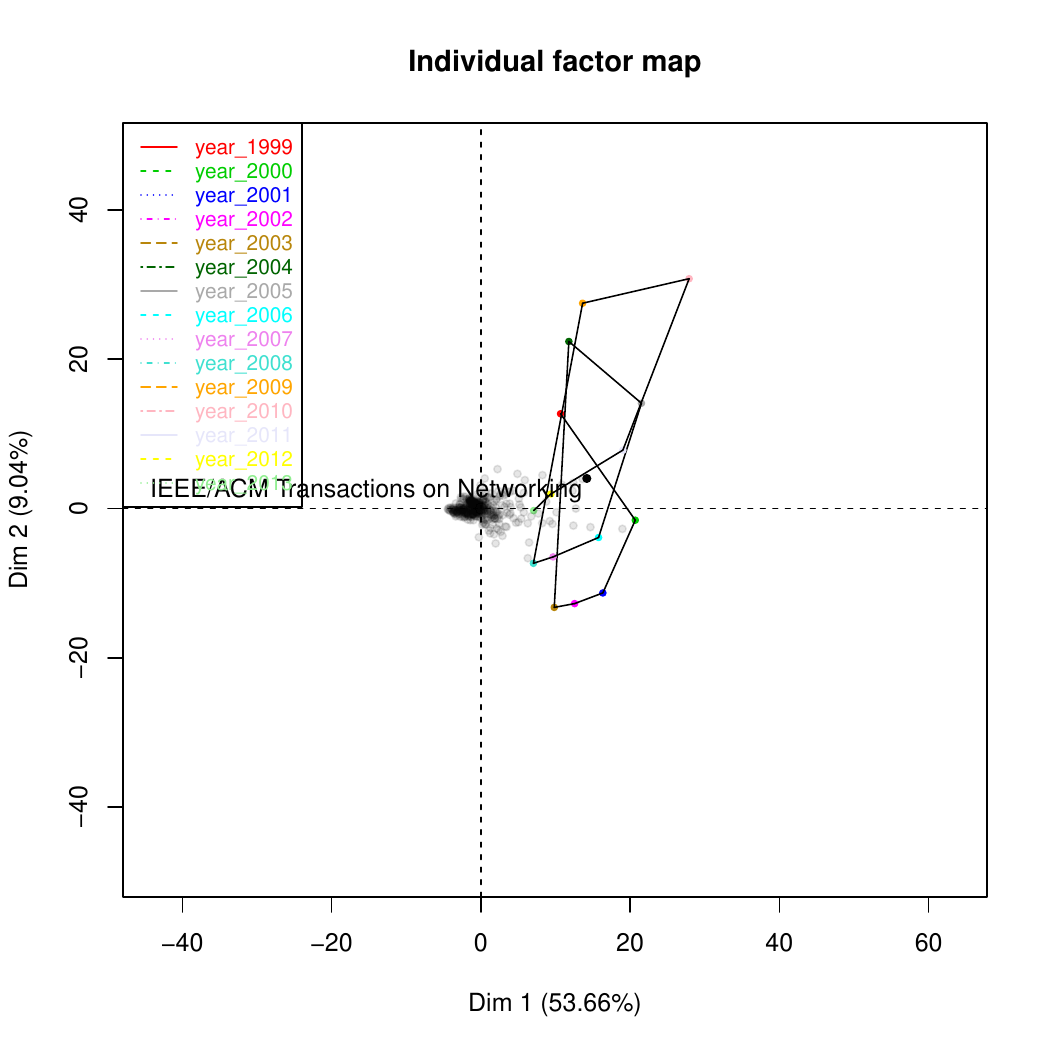} 
 \caption{MFA plots. Top: journals and metrics. Bottom: trajectories of IEEE and JSS  \label{fig:mfa}}
  \end{figure}

Figure \ref{fig:mfa} shows that journals on the right of the map take high values for all the metrics through the years (variables highly correlated to the first dimension on the metrics plot) and 
can be considered as  ``good" journals, whereas journals on the left take smaller values. In the upper part lie the journals with high scores for the SRJ measures meaning that they are prestigious.
``JSS'' is on the side of the good journals. Moreover, its trajectory has considerably improved through the years and is nowadays on the side of the best ones.
On the contrary, the trajectory of the ``IEEE/ACM Transactions on Networking" journal shows a slow decline since 2002.

\section{Conclusion}

We implemented several methods for low rank matrix estimation for complete data.
We also proposed development and implementation of some of these methods to handle incomplete data.
Although computationally intensive, these imputation methods are novel and promising.
% since the \textit{iterative  ATN algorithm} is performed for each cell of the data matrix to compute the divergence \eqref{eq:divna}.
%This later case is promising despite its high computational cost.
%Indeed, when selecting the two tuning parameters with SURE \eqref{eq:sureatnna} or GSURE, 
%the \textit{iterative  ATN algorithm} is performed for each cell of the data matrix to compute the divergence \eqref{eq:divna}.
A future version of the package will focus on improving the computational cost. 
In addition, GSURE would benefit from a better initial value, which is difficult to find in low signal-to-noise ratio. %In the current implementation of \pkg{denoiseR}, we use the package \pkg{parrallel} to use the different cores of the machine if available.  However, more work has to be done to know how to reduce the computational costs. We recommend using cluster computing if possible.
%Imputation with random forests also requires a lot of computer power. % since the method fits one random forest per variable and cycles through variables until convergence. %\citet{Audi15} noted that it takes 1.75 hours for a forest to impute a data with 6000 rows and 30 variables in comparison to 1 minute for an iterative SVD (without selection of any tuning parameters). 
%We should also warn the user to be careful with single imputation methods. Indeed, if they apply a statistical analysis on a completed data set, the variability of the estimators will be underestimated since it does not include the variability of the prediction.  Multiple imputation methods  \citep{Schafer97, Little02} can be considered to address this issue.  
%We finish by discussing some opportunities for further research.  There appear to be many potential applications for using ISA as a method to impute count data.  In addition, we would also like to define robust version of ATN with SURE with outliers cells.
A robust extension of our ATN method could also be useful when some cells may contain outliers.

%Note that the aim in matrix completion, is to predict as well as possible the missing values whereas in the missing %data litterature the aim is to perform a statistical analysis despite the missing entries, i.e. estimating parameters and %their variance as better as possible. In this latter case, single imputation is not necessarely a method to recommmand.
%\item estimate the noise variance with missing values, more research. 

%Centering?

%Imputation count data. However,  it could be interesting to assess it as an imputation method for count data as well.

%It is very common to visualize data with principal component analysis (PCA). PCA boils down to performing the truncated SVD of the data matrix $X$ as in %\eqref{eq:tsvd}. If we believe in model \eqref{mod:gauss}, it is better to use ATN for instance to recover the signal. Then, we can still represent the results  
%After each imputation step in the \textit{iterative ATN algorithm}, the means of the variables change. Consequently, it is necessary to recenter the data after %each imputation step. In the complete case, centering is often carried
%out prior to analysis and thus often regarded as a pre-processing step. In the incomplete case,
%it is necessary to consider the centering process as a part of the analysis. As far as we know,
%many algorithms performing SVD with missing values do not include these re-centering steps and thus may lead to poor results.

\section{Acknowledgement}

The authors thank Achim Zeiles from providing the impact factor data and Ludovic Lebart fo the presidents data. The authors are also very grateful to the reviewers and AE for their insightful comments. 

\bibliography{jss2909.bib}
\end{document}